\documentclass[floatfix,aps,prl,preprint,superscriptaddress]{revtex4}

\usepackage[pdftex]{graphicx}
\usepackage{times,amsmath,amssymb}

\usepackage{color}
\usepackage[colorlinks = true,
            linkcolor = blue,
            urlcolor  = blue,
            citecolor = blue,
            anchorcolor = blue]{hyperref}

\begin{document}

\title{Wide dynamic range charge sensor operation by high-speed feedback control of radio-frequency reflectometry}

\author{Yoshihiro Fujiwara}
\affiliation{Research Institute of Electrical Communication, Tohoku University, 2-1-1 Katahira, Aoba-ku, Sendai 980-8577, Japan}
\affiliation{Graduate School of Engineering, Tohoku University, 6-6 Aramaki Aza Aoba, Aoba-ku, Sendai 980-0845, Japan}

\author{Motoya Shinozaki}
\affiliation{WPI-Advanced Institute for Materials Research, Tohoku University, 2-1-1 Katahira, Aoba-ku, Sendai 980–8577, Japan}

\author{Kazuma Matsumura}
\affiliation{Research Institute of Electrical Communication, Tohoku University, 2-1-1 Katahira, Aoba-ku, Sendai 980-8577, Japan}
\affiliation{Graduate School of Engineering, Tohoku University, 6-6 Aramaki Aza Aoba, Aoba-ku, Sendai 980-0845, Japan}

\author{Kosuke Noro}
\affiliation{Research Institute of Electrical Communication, Tohoku University, 2-1-1 Katahira, Aoba-ku, Sendai 980-8577, Japan}
\affiliation{Graduate School of Engineering, Tohoku University, 6-6 Aramaki Aza Aoba, Aoba-ku, Sendai 980-0845, Japan}

\author{Riku Tataka}
\affiliation{Research Institute of Electrical Communication, Tohoku University, 2-1-1 Katahira, Aoba-ku, Sendai 980-8577, Japan}
\affiliation{Graduate School of Engineering, Tohoku University, 6-6 Aramaki Aza Aoba, Aoba-ku, Sendai 980-0845, Japan}

\author{Shoichi Sato}
\affiliation{Research Institute of Electrical Communication, Tohoku University, 2-1-1 Katahira, Aoba-ku, Sendai 980-8577, Japan}

\author{Takeshi Kumasaka}
\affiliation{Research Institute of Electrical Communication, Tohoku University, 2-1-1 Katahira, Aoba-ku, Sendai 980-8577, Japan}

\author{Tomohiro Otsuka}
\email[]{tomohiro.otsuka@tohoku.ac.jp}
\affiliation{WPI-Advanced Institute for Materials Research, Tohoku University, 2-1-1 Katahira, Aoba-ku, Sendai 980–8577, Japan}
\affiliation{Research Institute of Electrical Communication, Tohoku University, 2-1-1 Katahira, Aoba-ku, Sendai 980-8577, Japan}
\affiliation{Graduate School of Engineering, Tohoku University, 6-6 Aramaki Aza Aoba, Aoba-ku, Sendai 980-0845, Japan}
\affiliation{Center for Science and Innovation in Spintronics, Tohoku University, 2-1-1 Katahira, Aoba-ku, Sendai 980-8577, Japan}
\affiliation{Center for Emergent Matter Science, RIKEN, 2-1 Hirosawa, Wako, Saitama 351-0198, Japan}

\date{\today}

\begin{abstract}
Semiconductor quantum dots are useful for controlling and observing quantum states and can also be used as sensors for reading out quantum bits and exploring local electronic states in nanostructures.
However, challenges remain for the sensor applications, such as the trade-off between sensitivity and dynamic range and the issue of instability due to external disturbances.
In this study, we demonstrate proportional-integral-differential feedback control of the radio-frequency reflectometry in GaN nanodevices using a field-programmable gate array. 
This technique can maintain the operating point of the charge sensor with high sensitivity.
The system also realizes a wide dynamic range and high sensor sensitivity through the monitoring of the feedback signal.
This method has potential applications in exploring dynamics and instability of electronic and quantum states in nanostructures.
\end{abstract}

\maketitle
% \section{I. Introduction}
Semiconductor quantum dots have been widely studied due to their ability to artificially control and observe quantum states~\cite{tarucha1996shell, kouwenhoven1997excitation, kouwenhoven2001few}. 
They can also be used as charge sensors by coupling them to a target system, allowing observation of the quantum state of the target quantum dots~\cite{reilly2007fast, barthel2009rapid, yoneda2014fast}. 
This technique is also demonstrated by using quantum point contact and is useful for reading out quantum bits~\cite{field1993measurements, sprinzak2002charge, elzerman2003few}.
Furthermore, such sensors are useful for exploring local electronic states in nanostructures and are an important tool for investigating material properties~\cite{otsuka2015fast, otsuka2017higher, otsuka2019difference}.
It is always necessary to set the operating point of the charge sensor for high sensitivity, and there is an issue of instability of the operating point due to external disturbances, such as fluctuations of the charge states around the potential that forms the quantum dots~\cite{kirton1989noise, jung2004background, paladino20141}.

Feedback control that continuously monitors the state of the device is considered adequate to address this issue.
As an example of feedback control in quantum devices, the reduction of charge fluctuation in GaAs quantum dots has been reported using proportional-integral-differential (PID) feedback control~\cite{nakajima2021real}.
In addition to such feedback control, real-time processing of states in quantum dots has recently been demonstrated and is garnering attention for applications such as quantum bit operations~\cite{shulman2014suppressing, hornibrook2015cryogenic, conway2016fpga, kawakami2016gate, mills2019computer, nakajima2020coherence, zwolak2020autotuning, kanhirathingal2022feedback, xu2023noise, kobayashi2023feedback}.
In such cases, field-programmable gate arrays (FPGAs) are used because central processing units (CPUs) are slow and cannot sufficiently compensate for fast fluctuations of the states.
FPGAs can operate at significantly faster speeds and allow for flexible and immediate changes to digital signal processing circuits through hardware programming.
Therefore, they are useful in measurement systems that require flexible specifications. 
In order to advance the development of quantum information processing and sensors, it is important to construct measurement systems that combine the charge sensors with FPGAs.

The high-speed feedback control with FPGA is also useful to solve the existing challenges for the sensor applications of semiconductor quantum dots.
For example, the dynamic range of sensing is limited by the Coulomb peak width, resulting in a trade-off between sensor sensitivity and the dynamic range.
%Currently, the FPGA technique in quantum dots is still in its early stages of development and requires more study and reporting.
Here, we demonstrate the radio-frequency (rf) reflectometry in GaN nanodevices, which technique realizes high-speed readout to explore quantum dynamics~\cite{schoelkopf1998radio, reilly2007fast, barthel2009rapid}, and its PID feedback control is implemented by an FPGA.
We analyze the detailed behavior of the PID controllers for GaN nanodevices, including response speed and noise.
We also utilize the derivative term of the PID parameter enabling the fast feedback operation and show the fast response, which was not used in the previous report.
Analysis of the noise in the feedback signal reveals that it reflects the original noise behavior of the device, and we can detect the fluctuation by monitoring the feedback signal with keeping the optimal operation point.
From this insight, we demonstrate that the PID controller can achieve a wide dynamic range and high sensor sensitivity, which was previously a challenge.

% \section{II. Experimental setup}
% \subsection{A. Sample preparation and characteristics}
We treat GaN/AlGaN heterostructures, which exhibit high electron mobility in two-dimensional electron gases, making them attractive materials not only for electronics applications such as high electron mobility transistors~\cite{ambacher1999two, manfra2004electron, thillosen2006weak, shchepetilnikov2018electron}, but also from the perspective of quantum devices~\cite{chou2005high, chou2006single, ristic2005columnar, nakaoka2007coulomb}.
The wide and direct band gap in GaN offers the potential for the development of new quantum devices that operate at higher temperatures and can be coupled with light.
Despite the fact that GaN is attractive for quantum device applications, important techniques such as charge state readout using rf-reflectometry have not yet been reported.
Here, we demonstrate the rf-reflectometry in GaN nanodevices.

\begin{figure}
\begin{center}
\includegraphics{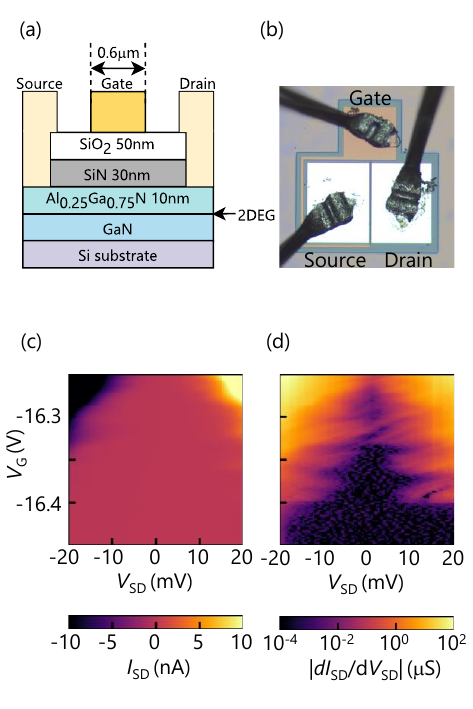}
\caption{(a) Schematic of the device structure. (b) Optical microscope image of the GaN device. (c) Gate voltage $V_{\rm G}$ and a source-drain bias voltage $V_{\rm SD}$ dependence of a source-drain current $I_{\rm SD}$, and (d) its numerical derivative. All experiments are conducted at 4.2~K.}
\label{fig1}
\end{center}
\end{figure}

The schematic of the device structure is shown in Fig.~\ref{fig1}(a) and (b).
A stack structure of GaN/Al$_{\rm 0.25}$Ga$_{\rm 0.75}$N (10nm)/SiN (30nm)/ SiO2 (50 nm) is grown by chemical vapor deposition on a silicon substrate.
The source and drain contacts are Ti/Al electrodes, and a gate is TiN with a length of 0.6 $\mathrm{\mu m}$.
The two-dimensional electron gas is formed at the interface between the GaN and AlGaN layers.
Fig.~\ref{fig1}(c) shows a gate voltage $V_{\rm G}$ and a source-drain bias voltage $V_{\rm SD}$ dependence of a source-drain current $I_{\rm SD}$.
We note that all measurements in this letter are carried out at a temperature of 4.2~K.
We focus on the near pinch-off region, and $I_{\rm SD}$ seems to be significantly suppressed.
This corresponds to Coulomb blockades of quantum dots formed by defects and/or impurities near the channel in the field effect transistor~\cite{otsuka2020formation, matsumura2023channel}.
Figure~\ref{fig1}(d) shows the numerical derivative of $I_{\rm SD}$.
Coulomb diamonds are observed, indicating the formation of quantum dots in our device.
The diamonds are not completely closed at $V_{\rm SD}=0$ in Fig~\ref{fig1}(a), indicating that multiple quantum dots are formed in this device.
% A possible origin of formation is the electrostatic potential fluctuation by defects and/or impurities near the channel.

% \subsection{B. Measurement setup of radio-frequency reflectometry}
\begin{figure}
\begin{center}
\includegraphics{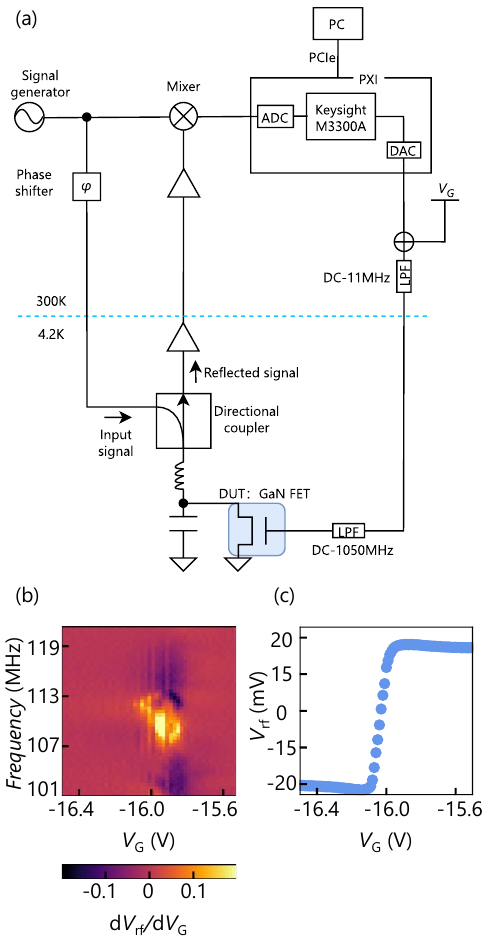}
\caption{(a) Measurement setup of radio-frequency (rf) reflectometry with the feedback sysytem. (b) Gate voltage $V_{\rm G}$ and frequency dependencies of ${\rm d}V_{\rm rf}/{\rm d}V_{\rm G}$. (c) Pinch-off behavior measured by the rf-reflectometry.}
\label{fig2}
\end{center}
\end{figure}
Broad measurement bandwidth is required for the feedback system to show enough performance.
In the case of the typical direct current measurement, such as we used to check the basic properties of the device, the bandwidth is limited by stray capacitances in the measurement setup.
In order to improve the bandwidth, rf-reflectometry is one of the powerful techniques.
This technique has been demonstrated in quantum point contacts and dots in gallium arsenide~\cite{reilly2007fast, barthel2009rapid}, Si/Ge~\cite{noiri2020radio}, and graphene devices~\cite{banszerus2021dispersive, johmen2022radio}.

We construct the rf-reflectometry setup with the feedback system as shown in Fig.~\ref{fig2}(a).
The input rf signal is applied to the resonator through the phase shifter and the directional coupler.
The resonator is constructed by an inductor $L=1.2 \, \mu \mathrm{H}$ and stray capacitance $C$ formed in the measurement board and the device.
The reflected rf signal is amplified and demodulated using the local signal, and the rectified voltage $V_{\rm rf}$ is sampled by a Keysight M3300A digitizer.
The sampling rate is 100 MS/s.
We check the resonator characteristics as shown in Fig.~\ref{fig2}(b), and there is a clear sensitive point of $V_{\rm rf}$ to $V_{\rm G}$.
To achieve maximum sensitivity, we set the rf frequency to 109.5 MHz and optimize the phase shift of the input rf signal.
% In this resonance circuit, we design that $L=1.2 \, \mu \mathrm{m}$ and $C=0.46 \, \rm{pF}$ which parameters promise a resonance frequency of about 214 MHz.
% This condition leads to the sensitivity to the device conductance $G \sim 20$ $\mu$S, which value is typically observed in quantum transport.
% However, the observed resonance frequency is much smaller than expected, and it is due to the large stray capacitance in our device.
% That capacitance is estimated to be 1.3 pF from the resonance frequency.
In this resonance circuit, the resonator is sensitive to the $G \sim 70$ $\mu$S, which satisfies the impedance matching condition.
By designing the device structure, it is possible to adjust the impedance matching condition to around $G \sim 20$ $\mu$S, which is observed in quantum dots.
% and observed Coulomb diamonds show the conductance far away from the sensitive region.
In this case, we focus on the simple pinch-off state showing the large conductance changes as shown in Fig.~\ref{fig2}(c).
This behavior is similar to that of a quantum point contact.
We set the measurement integration time at $10 \, \mu \rm s$ to reduce the noise in Fig.~\ref{fig2}(c).
An impedance-matching condition is satisfied at $V_{\rm G}=-16$ V where $V_{\rm rf}=0$.
The phase between the reflected and local signal is inverted here, resulting in the inversion of the sign of $V_{\rm rf}$.

\begin{figure*}
\begin{center}
  \includegraphics{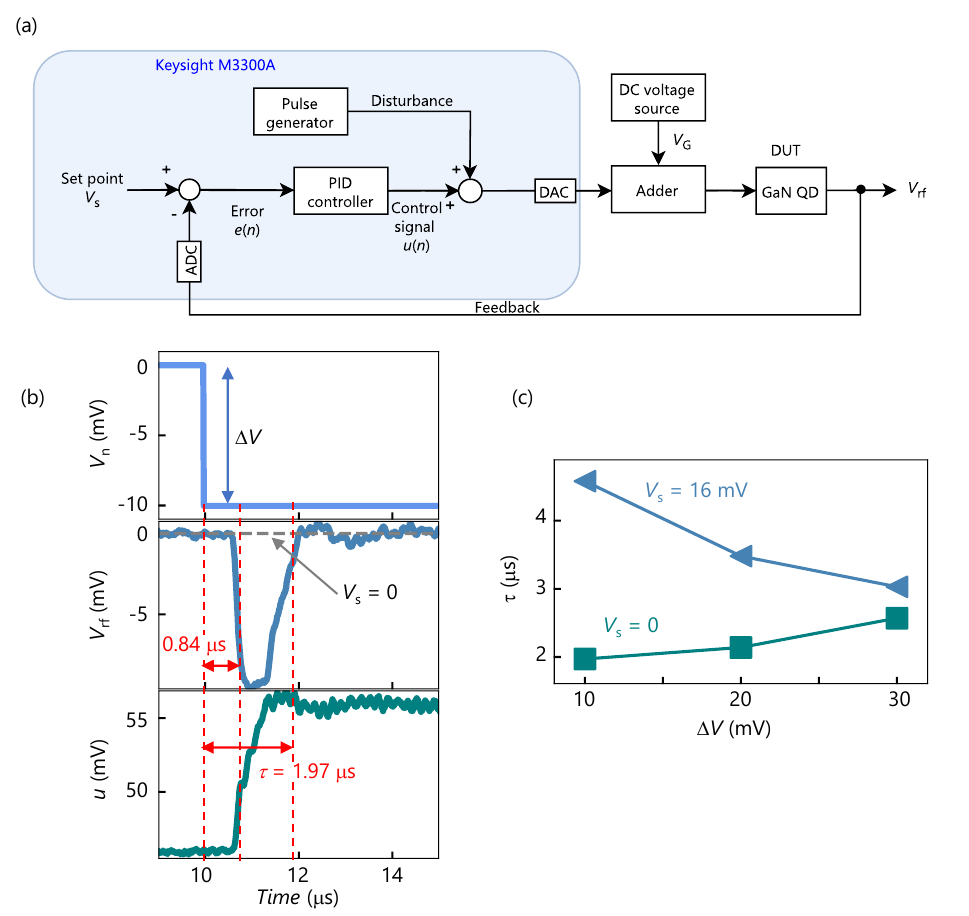}
  \caption{(a) Block diagram of our feedback system.
  (b) Time-traces of the synthesized step disturbance (top), monitored $V_{\rm rf}$ (middle), and the PID output $u$ (bottom).
  (c) Step disturbance amplitude dependence of $\tau$. We tuned the PID parameters at $\Delta V=10$ mV and $V_{\rm s}=0$.}
  \label{fig3}
\end{center}
\end{figure*}
% \section{III. Results}
% \subsection{A. Feedback operation by PID controller}

We describe our feedback system using the PID controller.
Figure~\ref{fig3}(a) shows the block diagram of the feedback loop.
The output of the PID controller (control signal $u(t)$) is expressed as

\begin{equation}
u(t) = K_{\rm P} e(t)+K_{\rm I} \int e(t) {\rm d}t + K_{\rm D} \frac{{\rm d}e(t)}{{\rm d}t},
\label{eq1}
\end{equation}
\begin{equation}
e(t) = V_{\rm s}-V_{\rm rf}(t),
\label{eq2}
\end{equation}
where, $t$, $V_{\rm s}$, $K_{\rm P}$, $K_{\rm I}$, and $K_{\rm D}$ are the continuous-time, the set point voltage, coefficients for the proportional, integral, and derivative terms, respectively.
In order to treat this system in the discrete-time $n$ domain, the Laplace transformation and the bilinear transformation are aaplied to Eq.~\ref{eq1}.
The transfer function of the PID controller $H(z)$ is described as
\begin{equation}
H(z) = G_{\rm P} + G_{\rm I} \frac{1+z^{-1}}{1-z^{-1}} + G_{\rm D} \frac{1-z^{-1}}{1+D z^{-1}},
\label{eq3}
\end{equation}
where, $G_{\rm P} = K_{\rm P}$, $G_{\rm I} = K_{\rm I} \frac{T_{\rm S}}{2}$, $G_{\rm D}=\frac{2K_{\rm D}}{T_{\rm S}+2T_{\rm F}}$, $D=\frac{T_{\rm S}-2T_{\rm F}}{T_{\rm S}+2T_{\rm F}}$, $T_{\rm S}$ is the sampling time, and $T_{\rm F}$ is the time constant of a differentiator, respectively.
From $H(z)$, the output of the PID controller in $n$ domain $u(n)$ can be obtained as
\begin{equation}
u(n) = u_{\rm P}(n) + u_{\rm I}(n) + u_{\rm D}(n) + u(n-1),
\label{eq4}
\end{equation}
\begin{equation}
u_{\rm P}(n) = G_{\rm P} e(n),
\label{eq5}
\end{equation}
\begin{equation}
u_{\rm I}(n) = G_{\rm I} \left[e(n)+e(n-1)\right] + u_{\rm I}(n-1),
\label{eq6}
\end{equation}
\begin{equation}
u_{\rm D}(n) = G_{\rm D} \left[e(n)-e(n-1)\right] - Du_{\rm D}(n-1),
\label{eq7}
\end{equation}
\begin{equation}
e(n) = V_{\rm s}-V_{\rm rf}(n).
\label{eq8}
\end{equation}
The proportional, integral, and derivative terms are processed in parallel.
In addition, we use the digital low-pass filter and averaging process before the input of the PID controller in order to reduce the aliasing.
The fastest $I/O$ latency of the M3300A is 100 ns.

Figure~\ref{fig3}(b) shows the response of our feedback system.
A synthesized step disturbance $V_{\rm n}$ is applied to $V_{\rm G}$, where $V_{\rm G}=-16.03$ V and $V_{\rm s}=0$.
This is intended for electrostatic potential fluctuation by defects and/or impurities near the channel, which acts as the effective gate voltage.
Then, the PID controller operates to compensate for $V_{\rm n}$, resulting in the stabiliztion of $V_{\rm rf}$.
Here, we set the PID parameters $G_{\rm P}=0.10$, $G_{\rm I}=0$, $G_{\rm D}=0.65$, $D = 0.80$, $T_{\rm F}=5$ ns and $T_{\rm S}=90$ ns, respectively.
These time traces are averaged by 1000 trials.
When $V_{\rm G}$ is shifted by $V_{\rm n}$, $V_{\rm rf}$ is rapidly returned and stabilized at $V_{\rm s}$ by the PID output $u(n)$.
The PID controller operates successfully.
One of the differences from the previous report is the setting of parameters.
We use mainly proportional and derivative terms, while the previous study adapted proportional and integral terms~\cite{nakajima2021real}.
The derivative term suppresses an overshoot induced by $K_{\rm P}$ and contributes to the fast response.
% In the case of our system, the $K_{\rm I}>6.1$ is imposed by the $T_{\rm S}=100$ ns and a resolution of $G_{\rm I}$.
% This value is quite large.
% Therefore, the $u$ oscillates by the integral term when the PID controller has a high-speed clock of over 10 MHz.
In the case of our system, we do not use the integral term because $u$ oscillates by the term when the PID controller has a high-speed clock of over 10 MHz.
We define the fall time as the time between $V_{\rm n}$ input timing and the time for $V_{\rm rf}$ to reach 90\% of the maximum changes, and the rise time $\tau$ as the time to return 10\% of the maximum changes.
The fall-time is mainly due to the system $I/O$ latency.
Under this condition, we achieve $\tau = 1.97~\mu$s, faster than in the previous study.
Figure~\ref{fig3}(c) shows the disturbance amplitude $\Delta V$ dependence of $\tau$ at $V_{\rm s}=0$ and $16$ mV.
We use the same PID parameters in all measurement conditions, and the feedback control is nicely operated in micro-second order even the condition is changed.
$\tau$ increases with increasing $\Delta V$ for $V_{\rm s}=0$ while it decreases for $V_{\rm s} = 16$ mV.
Because the PID parameters are optimized at $V_{\rm s}=0$ and $\Delta V=10$ mV, $\tau$ increases when the condition changes away from this.
As demonstrated, the feedback control might not always perform optimally, indicating that it would be better to choose optimum parameters for each $V_{\rm s}$ in order to improve performance.
Our PID controller is easy to tune the parameter because of the use of the proportional and derivative processes.
    
% \subsection{B. Analysis of the power spectral density}
\begin{figure}
\begin{center}
  \includegraphics[width=8cm]{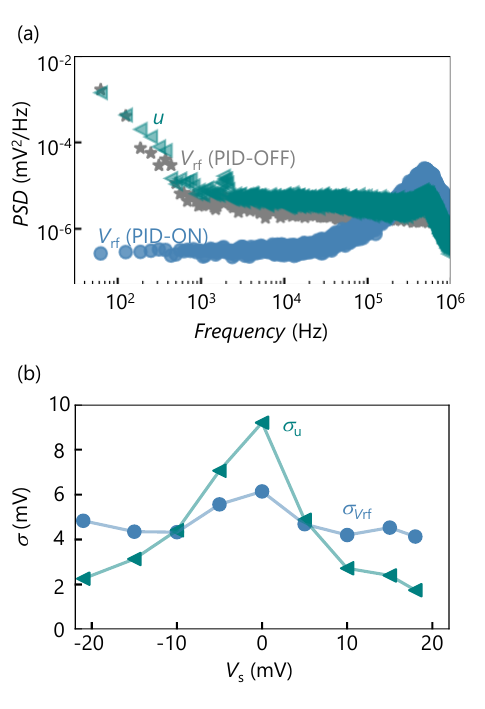}
  \caption{(a) Noise PSDs of $V_{\rm rf}$ with PID on/off and $u$ at $V_{\rm s}=0$. 
  (b) $V_{\rm s}$ dependence of $\sigma_{u}$ and $\sigma_{V{\rm rf}}$.}
  \label{fig4}
\end{center}
\end{figure}
We also investigate the noise power spectral density (PSD) of $V_{\rm rf}$ with the PID controller and its output $u$.
Figure~\ref{fig4}(a) shows the PSD of $V_{\rm rf}$ with the PID on/off and $u$ at $V_{\rm s}=0$.
The bandwidth is limited by a digital low-pass filter of 5 MHz.
At a glance, the low-frequency noise including the flicker noise below 100 kHz is significantly suppressed by the PID control.
This cut-off frequency of the PID controller corresponds to $1/\tau$ we observed in Fig.~\ref{fig3}.
We also focus on the PSD of $u$, and it shows a similar behavior of $V_{\rm rf}$ with the PID off, which means that we can probe and track the noise information slower than the feedback control by monitoring $u$ with keeping $V_{\rm rf}$ stabilized.

Figure~\ref{fig4}(b) shows the readout deviation $\sigma$ of $V_{\rm rf}$ and $u$.
While $\sigma_{V{\rm rf}}$ is almost constant by changing $V_{\rm s}$, $\sigma_{u}$ depends on $V_{\rm s}$.
The flicker noise contributes to $\sigma_{u}$ and is proportional to $|\frac{{\rm d}V_{\rm rf}}{{\rm d}V_{\rm G}}|$.
Therefore, the $V_{\rm s}$ dependence of $\sigma_{u}$ reflects $|\frac{{\rm d}V_{\rm rf}}{{\rm d}V_{\rm G}}|$.
Note that the PID parameters are optimized at $V_{\rm s}=0$, and its response deteriorates away from this optimal point as shown in Fig.~\ref{fig3}(c), which may also affect $\sigma_{u}$.
This behavior agrees well with previous findings~\cite{jung2004background, shinozaki2021gate}.
On the other hand, the background of $\sigma_{V{\rm rf}}$ is larger than that of $\sigma_{u}$.
The PSD of $V_{\rm rf}$ exhibits a slight increase in a frequency domain higher than the feedback control frequency of 100 kHz, which appears to increase $\sigma_{V{\rm rf}}$ independent of $V_{\rm s}$.

% \subsection{C. Sensor operation}
\begin{figure}
\begin{center}
  \includegraphics{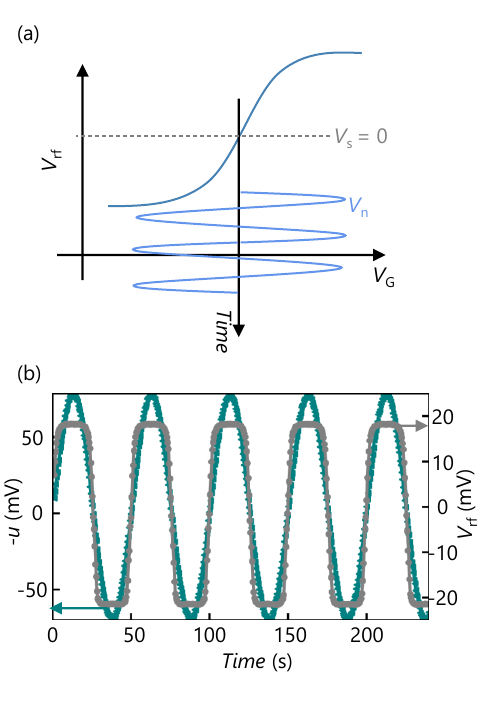}
  \caption{(a) Schematic of the measurement for charge sensing.
  (b) Real-time trace of $V_{\rm rf}$ with the PID off and $u$ with $V_{\rm n}$. A sinusoidal fluctuation is applied to the $V_{\rm G}$.}
  \label{fig5}
\end{center}
\end{figure}

Finally, we demonstrate the wide-range charge sensing by the PID controller.
Quantum dots can be utilized as charge detection sensors, and they can probe local electronic states in nanostructures~\cite{otsuka2015fast, otsuka2017higher, otsuka2019difference}.
Due to the electrostatic coupling between the quantum dots and local electronic states, the changes in the charge states act as effective $V_{\rm G}$.
For the charge sensor application, the sensor state should be stabilized.
In addition, the saturation region of $V_{\rm G}$ dependence of the $V_{\rm rf}$ has no sensitivity to the charge state, indicating that the available sensing range is limited by the width of the region that has a slope of $V_{\rm rf}$ to $V_{\rm G}$.
While a large electrostatic coupling leads to a large signal that detects the charge state, the signal could exceed the sensitive region of $V_{\rm rf}$.
The PID controller solves these concerns by monitoring the signal through the feedback control.
We applied the sinusoidal signal as the simulated charge state change, where we set the $V_{\rm s}=0$.
The schematic of the measurement is shown in Fig.~\ref{fig5}(a).
At first, we monitor $V_{\rm rf}$ with PID off as shown in Fig.~\ref{fig5}(b).
It can be seen that the signal saturates around $V_{\rm rf} \sim \pm 20$ mV, and we cannot probe any charge states in these regions.
Next, we monitor $u$ with the PID on.
Clearly seen in the result, the charge state is perfectly tracked by monitoring $u$ with keeping the operation point stabilized.
The PID controller serves as a sensitive and wide-dynamic-range probe for the local electronic states in nanostructures.
A tracking range is limited by the maximum range of $u$, which is 1.5 V.
This value is greater than the range of detection with the PID off.
The PID controller is expected not only to stabilize the quantum dots but also to be a high-performance probing tool for local electronic states.

% \section{IV. Conclusion}
% In conclusion, we demonstrate the PID feedback control of rf-reflectometry in GaN nano-device by FPGA.
% We observe the formation of the quantum dots in GaN at 4K near the pinch-off region.
% The pinch-off state can be measured by rf-reflectometry, and we show the PID feedback control with a bandwidth of 100 kHz.
% The noise PSDs of the $V_{\rm rf}$ with PID on/off and $u$ are evaluated.
% The PID control significantly suppresses the low-frequency noise, including the flicker noise below 100 kHz, and PSDs of $V_{\rm rf}$ with the PID off and $u$ are similar behavior.
% We also show the charge sensor operation exceeded the conventional system by monitoring the $u$.
% This method satisfies both device stability and wide sensing range with high-sensitivity.
% The present system and operation are useful for the application of quantum dots and exploring the instability quantum state in nanostructures for many materials.

In conclusion, we demonstrate the rf-reflectometry in GaN nanodevices, which enables high-speed readout for investigating quantum dynamics. 
We implement PID feedback control by FPGAs and analyze the behavior of the PID controller, including response speed and noise PSD. 
By utilizing the derivative term in the PID, we achieve the fast response. 
The PID control significantly suppresses low-frequency noise, including flicker noise below 100 kHz, and PSDs of $V_{\rm rf}$ with the PID off and $u$ show similar behavior.
This result allows us to detect the change in charge states by monitoring $u$ while keeping the optimal operating point. 
Consequently, we demonstrate that PID controllers successfully enable both a wide dynamic range and high sensor sensitivity.
The present system and operation are useful for applications in quantum dots and exploration of the electronic and quantum states in nanostructures.

% \section{Acknowledgements}
The authors thank N, Ito, T, Tanaka, K, Nakahara, and RIEC Fundamental Technology Center and the Laboratory for Nanoelectronics and
Spintronics for fruitful discussions and technical support. 
Part of this work is supported by MEXT Leading Initiative for Excellent Young Researchers, 
Grants-in-Aid for Scientific Research (21K18592, 23H01789, 23H04490),
Rohm Collaboration Project,
Fujikura Foundation Research Grant, 
Tanigawa Foundation Research Grant, 
Maekawa Foundation Research Grant, 
The Foundation for Technology Promotion of Electronic Circuit Board,
Iketani Science and Technology Foundation Research Grant,
and FRiD Tohoku University.

% \begin{references}
\bibliography{reference.bib}
% \end{references}
\end{document}